# A Convergent Method for Calculating the Properties of Many Interacting Electrons

Roger Haydock

Department of Physics and Materials Science Institute

University of Oregon, Eugene OR 97403-1274, USA

**Abstract**

A method is presented for calculating binding energies and other properties of extended interacting systems using the projected density of transitions (PDoT) which is the probability distribution for transitions of different energies induced by a given localized operator, the operator on which the transitions are projected. It is shown that the transition contributing to the PDoT at each energy is the one which disturbs the system least, and so, by projecting on appropriate operators, the binding energies of equilibrium electronic states and the energies of their elementary excitations can be calculated. The PDoT may be expanded as a continued fraction by the recursion method, and as in other cases the continued fraction converges exponentially with the number of arithmetic operations, independent of the size of the system, in contrast to other numerical methods for which the number of operations increases with system size to maintain a given accuracy. These properties are illustrated with a calculation of the binding energies and zone-boundary spin-wave energies for an infinite spin-1/2 Heisenberg chain, which is compared with analytic results for this system and extrapolations from finite rings of spins.

1. Interacting Electrons

Calculating binding energies for even a few interacting electrons is a formidable problem. Because of the interactions, the Schroedinger equation cannot be solved analytically, nor does it separate, and the number of parameters needed for a variational solution grows exponentially with the number of electrons. The ground state energy for a pair of interacting electrons which hop between just two spatial orbitals is already a substantial calculation. More electrons or orbitals requires a computer, and fewer than 100 interacting electrons defeat any numerical attempt to solve the Schroedinger equation variationally.

The purpose of this paper is to show how properties of interacting electrons can be calculated so that the analytic or numerical effort required to obtain a given accuracy is independent of the number of electrons, and only grows logarithmically with the accuracy. This method arises from previous work on the states of non-interacting electrons in non-crystalline solids [1, 2, 3], and recent work [4] on the calculation of the projected density of transitions (PDoT) for interacting electrons. The main new result of this paper is that the binding energies and excitation energies of interacting electrons can be extracted from the PDoT, and this is illustrated with calculations for an infinite, spin-1/2 Heisenberg chain.

While the Hamiltonian for a Heisenberg chain is simple, it still possesses all the complexity of interacting systems. For example the Hamiltonian for a chain of N+1 electrons has dimension $2^{N+1}$, one for each configuration of the spins, and so has $2^{N+1}$ stationary states whose energies per bond must lie between the ferromagnetic and anti-ferromagnetic limits. The number of states increases exponentially with the number of electrons while the energies of the states only increase linearly, so the density of states must also increase exponentially with the number of electrons. Even taking into account that the interaction conserves total spin, which can at most increase linearly with the number of electron, the density of states for a given total spin still increases exponentially with the number of electrons.

In applying linear or non-linear variational methods to interacting systems, there must be a parameter for each degree of freedom which is to be varied independently, and so the number of variational parameters also increases exponentially with system size. Perturbative approaches diverge because the energy denominators decrease exponentially with the exponentially increasing number of coupled states in any energy interval. Other methods sample the states either statistically as in Monte Carlo, or by other criteria as in renormalization methods. The accuracy of Monte Carlo calculations only increases as the reciprocal of the root of the computational effort, and the errors in renormalization calculations depend on the criteria for neglecting states in a way which cannot be predicted in advance.

The method presented here has its physics roots in the black body theorem of von Laue

[5], which explains why the local (at a point) electromagnetic power spectrum is exponentially insensitive to its surroundings. Friedel [6] applied this idea to non-interacting electrons where the projected density of states (PDoS) - the total density of state weighted by the probability of finding an electron in a particular orbital - is exponentially insensitive to parts of the system distant from the orbital on which the states are projected. Friedel observed that moments of the PDoS could be calculated easily from powers of the electronic Hamiltonian, leading to local approaches to the electronic structure of solids [1] and to the recursion method for calculating projected densities of states from a continued fraction obtained by tridiagonalizing the electronic Hamiltonian. The focus on the PDoT and the use of energy-independent inner products, both of which are crucial for this approach, distinguish it from the projection methods developed by Zwanzig [7] and Mori [8] for correlation functions in classical and quantum statistical mechanics.

While the stationary states of interacting systems have energies which grow with the size of the system, the stationary transitions, operators transforming one stationary state into another, need not have energies which depend on the system's size because they are the differences between energies of stationary states. The transitions satisfy Heisenberg's equation [4] which is equivalent to Schroedinger's equation, but does not become singular as the system gets larger, again because the transition energies do not increase. Just as non-interacting states are projected on a localized orbital for the PDoS, stationary transitions are projected on a localized operator for the PDoT which is then the total density of transitions weighted by the probability that each transition is induced by the localized operator on which it is being projected. It is crucial to the work presented here to avoid the singular behavior of stationary states with increasing system size.

The mathematical roots of this work lie in the classical moment problem [9] and the properties of orthogonal polynomials [10]. The moment problem is that of reconstructing a distribution from its moments, integrals of the distribution over powers of the variable; in this work the distribution is the probability density for transitions and the variable is the energy of the transition. The surprising solution to the moment problem is that a continued fraction expansion of the distribution converges exponentially with the number of moments. Orthogonal polynomials in energy arise in this work as coefficients in expansions of the stationary transitions, and they are orthogonal with respect to integration over the PDoT. The numerical applications of these expansions are infinite dimensional analogs of the Lanczos method for finite matrices, in particular, convergence of the continued fraction expansion for the PDoS [11] corresponds to Paige's theorem [12] for the convergence of the Lanczos method.

The rest of this paper is organized into 5 further Secs. In the next Sec. the local properties of the PDoT and its calculation by the recursion method are reviewed briefly. In addition to the review, there is a construction of density matrices for stationary transitions from which expectation values can be calculated. In Sec. 3, the property of the PDoT that the transition which contributes at each energy is the one which is qualitatively most localized is used to show how binding energies, excitation energies, and other quantities can be calculated for individual equilibrium states. The way transitions with different localization properties contribute to the PDoT is illustrated in Sec. 4 which contains analytic examples of a localized transition which is degenerate with a band of delocalized transitions, and an example of two degenerate bands of transitions with different localization properties. In Sec. 5 is the description of a calculation of the equilibrium binding energies and zone boundary spin-wave energies for electrons in an infinite Heisenberg chain, which is compared to analytic results for this system and some numerical results for finite rings. The last Sec. contains a discussion of how this method is related to thermodynamical and statistical mechanical approaches.

While the Heisenberg chain is used throughout this paper to illustrate the method presented, it is intended to be clear that this method applies to a wide range of systems.

2. Projected Density of Transitions, the Recursion Method, and Expectation Values

This work is based on the two local properties of the PDoT analogous to properties of the PDoS: that it is exponentially insensitive to distant parts of the interacting system, and that the transition which contributes to the PDoT at each energy is the one which is most localized, and so changes the system least. Because of this, there follows a brief review of the relation of the PDoT to the energy resolvent for the evolution of operators, its calculation by the recursion method, and an expansion of the density matrices for stationary transitions.

For a system with N non-degenerate (to avoid complications) stationary energies $\{E_\alpha\}$ and corresponding states $\{\psi_\alpha(\mathbf{r})\}$, expressed as functions of $\mathbf{r}$ which stands for the dynamical variables such as position, spin, and so on. If u is an operator on which the transitions are to be projected, for example the annihilation operator for an electron in a particular orbital, then the PDoT for this system is defined to be,

$$g_u(E) = (1/N) \sum_{\alpha,\beta} \left| \int \psi_\beta(\mathbf{r})^* u \psi_\alpha(\mathbf{r}) d\mathbf{r} \right|^2 \delta(E - E_\beta + E_\alpha) \tag{1}$$

where the integral is over all values of the dynamical variables, and the sum is over all $N^2$ pairs of initial states $\psi_\alpha(\mathbf{r})$ and final states $\psi_\beta(\mathbf{r})$. Each choice of initial and final states contributes to the PDoT at an energy which is that of the final state minus that of the initial state, with a weight which is the squared magnitude of the matrix element of u between the initial and final states (of order 1/N from normalization of the states), divided by N (to normalize the integral of $g_u(E)$ over energy). For such a finite system, the PDoT is a weighted sum of delta-distributions and is degenerate at least to the extent that the transitions from each state to itself contribute with zero transition energy. The degeneracy of the PDoT can be greater if more than one pair of distinct initial and final states differ by the same energy.

      The PDoT is actually a generalization of the PDoS for a non-interacting system coupled to a particle bath (a source or sink of particles at some potential), as can be seen when u, in Eq. 1, is taken to be the annihilation operator for a particle in the single-particle orbital on which the states are to be projected. For a non-interacting system the change in the energy of the system due to removing a particle is independent of the state of the other particles. For a system of variable particle number, the initial states in Eq. 1 include every symmetrized product of every number of independent particle states, so the energy differences are simply the energies of the independent-particle states relative to the potential of the bath, and the weights are just the squared magnitudes of the overlaps between each independent-particle state and the orbital in which u annihilates a particle. This is just the PDoS for the independent particle states projected onto the orbital in which u annihilates an electron.

2.1 The Recursion Method

      The PDoT is $1/\pi$ times the magnitude of the imaginary part of the projected resolvent or Greenian[4],

$$R_u(E) = <u, (E - \mathbf{L})^{-1} u>, \qquad (2)$$

where $\mathbf{L}$ is the Liouvillian 'superoperator' which acts on an operator x by commuting the Hamiltonian with it,

$$\mathbf{L} x = [H, x]. \qquad (3)$$

The inner product in Eq. 2 does not affect the energies of transitions contributing to the PDoT, just the weights with which they contribute, so it may be chosen for convenience. Numerical methods are much more stable if $\mathbf{L}$ is Hermitian with respect to the inner product, and an example of such an inner product is the normalized trace,

$$\langle u, v \rangle = \sum_\alpha \int \phi_\alpha(\mathbf{r})^* \{u\dagger v\} \phi_\alpha(\mathbf{r}) d\mathbf{r} / \left[ \sum_\alpha \int \phi_\alpha(\mathbf{r})^* \phi_\alpha(\mathbf{r}) d\mathbf{r} \right] \qquad (4)$$

where $\{\phi_\alpha(\mathbf{r})\}$ is a complete orthonormal set of states, usually Slater determinants of localized orbitals. This has the added advantage that traces in the numerator and denominator of Eq. 4 need only include the orbitals whose occupations are changed by u or v; the rest cancel [4].

An efficient and accurate way of calculating the projected resolvent is with a continued fraction expansion [4]

$$R_u(E) = b_0^2 / (E - a_0 - b_1^2 / (E - a_1 - ... - b_n^2 / (E - a_n - ...) ...)), \qquad (5)$$

where the parameters $\{a_n\}$, and $\{b_n\}$ are tridiagonal matrix elements of **L** obtained by defining $u_{-1}$ to be zero, normalizing u with respect to the inner product in Eq. 4 to get $u_0$, and then constructing a sequence $\{u_n\}$ of operators which are orthonormal with respect to the inner product,

$$\langle u_n, u_m \rangle = \delta_{n,m}, \qquad (6)$$

and for which,

$$\mathbf{L}\, u_n = a_n u_n + b_{n+1} u_{n+1} + b_n u_{n-1}. \qquad (7)$$

Explicit formulas for the parameters and basis operators are:

$$a_n = \langle u_n, \mathbf{L}\, u_n \rangle, \qquad (8)$$

$$b_{n+1} = \sqrt{\langle (\mathbf{L} - a_n) u_n - b_n u_{n-1}, (\mathbf{L} - a_n) u_n - b_n u_{n-1} \rangle}, \qquad (9)$$

and,

$$u_{n+1} = [(\mathbf{L} - a_n) u_n - b_n u_{n-1}] / b_{n+1}. \qquad (10)$$

Section 4 has two examples where **L** is tridiagonalized analytically, while Sec. 5 illustrates numerical tridiagonalization. In other cases[2] an analytic tridiagonalization may be known for an approximate **L**, from which the corrections to $\{u_n\}$, $\{a_n\}$, and $\{b_n\}$ can be calculated perturbatively.

As in the case of states [1], the stationary or eigentransitions contributing to the PDoT can be constructed from the tridiagonalization of **L** using orthogonal polynomials $\{P_n(E)\}$ which are solutions to the same recurrence as the $\{u_n\}$

$$P_{n+1}(E) = [(E - a_n) P_n(E) - b_n P_{n-1}(E)] / b_{n+1}, \qquad (11)$$

but with the initial conditions that $P_{-1}(E)$ is zero, and $P_0(E)$ is 1. Using Eqs. 7 and 11, the transition,

$$\Psi_\alpha = \sum_{n=0}^{\infty} P_n(E_\alpha)\, u_n, \tag{12}$$

is a solution to,

$$\mathbf{L}\, \Psi_\alpha = E_\alpha\, \Psi_\alpha. \tag{13}$$

The relative weight of the transition at energy E in the first N elements of the tridiagonal basis is given by its Christoffel function [13],

$$w_N(E) = 1/\left[\sum_{n=0}^{N-1} P_n(E)^2\right], \tag{14}$$

which normalizes the transitions.

2.2 Density Matrices and their Expectation Values

These polynomial expansions of transitions contributing to the PDoT can be used to construct density matrices for the initial and final states in each transition. Because the Liouvillian commutes the Hamiltonian with operators, a product of stationary transitions is also a stationary transition. In particular the two products of a stationary transition $\Psi_\alpha$ and its Hermitian adjoint $\Psi_\alpha^\dagger$,

$$\rho_i(\alpha) = \Psi_\alpha^\dagger\, \Psi_\alpha, \text{ and } \rho_f(\alpha) = \Psi_\alpha\, \Psi_\alpha^\dagger, \tag{15}$$

are constant operators or constant density matrices because $\mathbf{L}\rho_i(\alpha)$ and $\mathbf{L}\rho_f(\alpha)$ are both zero, $\rho_i(\alpha)$ and $\rho_f(\alpha)$ are both Hermitian, and neither can have negative eigenvalues. The weights of various stationary states in $\rho_i(\alpha)$ and $\rho_f(\alpha)$ are discussed in the next Sec. so what follows is simply the mechanics of calculating expectation values for these density matrices.

The orthogonal polynomial expansion for the stationary transition $\Psi_\alpha$ contributing to the PDoT at energy $E_\alpha$, Eq. 12 is now used to obtain expectation values for the above density matrices. Given an operator Q which is finite ranged in the sense that it only changes the occupations of a finite number of localized orbitals, the expectation value of Q for $\rho_i(\alpha)$ is,

$$<<Q\ \rho_i(\alpha)>> = \lim_{N\to\infty} w_N(E_\alpha) \sum_{n=0}^{N} \sum_{m=0}^{N} P_n(E_\alpha)\ P_m(E_\alpha) <u_n, Q\ u_m>, \qquad (16)$$

where the Nth Christoffel function evaluated at $E_\alpha$, $w_N(E_\alpha)$, normalizes the expectation value in the same way as dividing by the partition function. A similar expression gives the expectation value of Q for $\rho_f(\alpha)$.

The above expression for the expectation value of Q contains a limit whose convergence properties can be seen more easily as the limit of expectation values of Q within the N+1-dimensional subspace of operators, spanned by $u_0, u_1, ..., u_{N-1}, u_N$. Viewing Q as a linear mapping of operators to operators, its finite range means that Q has a finite magnitude which bounds its magnitude within any finite subspace of operators. Consequently, the limit in Eq. 16 converges absolutely.

3. Weights of Transitions and States

The physical interpretation of formulas such as Eq. 5 for the PDoT, Eq. 12 for the stationary transitions, and Eq. 16 for expectation values depends on the weights with which different transitions and stationary states of the system contribute to these quantities. One stationary transition from each subspace of degenerate stationary transitions contributes to the PDoT [4] with a weight equal to the squared magnitude of its overlap with the projecting operator. To determine the weights of different states in the density matrices defined in Eq. 15, it is important to distinguish between pure transitions, those with a unique pair of initial and final states, and stationary transitions which can be superpositions of pure transitions which have the same energy but have different initial and final states. If there is degeneracy in the transitions with some energy $E_\alpha$, then the polynomial expansion for $E_\alpha$ can produce a stationary transition which is not pure, but is a superposition of pure transitions, and the resulting density matrices $\rho_i(\alpha)$ or $\rho_f(\alpha)$ will be weighted combinations of projection operators for the various initial or final states. However, if the transitions are non-degenerate, then the polynomial expansion has unique initial and final states, and each density matrix is a projection onto just one of these states.

3.1 Qualitative Properties of Transitions

The weight of a stationary transition in the PDoT is its Christoffel function, Eq. 14, which is the reciprocal of its normalization. This relation between normalization and weight also applies to pure transitions when expanded as a sum of operators in which the projecting operator $u_0$ has unit coefficient. Indeed, the smaller the component of $u_0$ is relative to other components in each pure transition, the larger the normalization, and hence the smaller the weight. Although the coefficients in the polynomial expansions are always finite, in infinite systems, it is possible for degenerate pure transitions to have divergent normalizations, a ratio of normalizations that is zero or infinity, in which case the relation between weight and normalization means that only the pure transition with the smaller normalization contributes to the PDoT and polynomial expansion while the one with the larger normalization is completely absent from both quantities. Analytic examples of degenerate transitions with divergent normalizations are given in Sec. 4.

The conclusion of the above argument is that if more than one pure transition has the same transition energy, any whose normalization (as defined above) is infinitely larger than any other does not contribute to the PDoT and has no component in the polynomial expansion for the stationary transition at that energy. Put in another way, only the pure transition with the smallest normalization, and any other pure transitions whose normalizations are only finite multiples of the smallest, contribute to the PDoT or the polynomial expansion. In infinite systems, an infinite number of qualitatively different normalizations are possible, for example normalizations which vary with the system size as different positive powers of the volume. For this reason, large parts of each subspace of degenerate transitions do not contribute to either the PDoT or polynomial expansions, and the initial and final states of these pure transitions have zero weight in the density matrices.

It is now argued that the qualitative differences in the normalizations of pure transitions reflect a more general qualitative distinction between transitions. Because the polynomials are finite, the components of a transition are always finite, and so a divergence in the normalizations of two transitions means that one transition contains infinitely more finite-ranged operators than the other. This may be interpreted in two ways: either that one transition is spread over an infinitely larger part of the system than the other, or that there is no mapping from one transition to the other which preserves the finite range of its components. Either interpretation of qualitative differences in the normalization implies a qualitative physical distinction.

Since degeneracy of qualitatively different pure transitions leaves the density matrices $\rho_i(\alpha)$ and $\rho_f(\alpha)$ as projections on the unique initial or final state of the pure transition with the qualitatively smallest normalization, it is only degeneracy of qualitatively similar transitions which remains to be considered. Such transitions must still have quantitative differences which are reflected in different quantum numbers as for example the degenerate transitions in a system of non-interacting particles where the degenerate pure transitions of one particle are distinguished by different quantum numbers of the other particles. If there are degenerate pure transitions which don't change some of the quantum numbers, then the polynomial expansion of a stationary transition at this energy also leaves the same quantum numbers unchanged, or, in other words, acts as the identity on the degrees of freedom described by those quantum numbers. The resulting density matrices are also identity operators for the degrees of freedom which do not affect the energy of the transition, and project only in the degrees of freedom which determine the energy of the transition. For the example of independent particles, out of all the many-particle states of the system, the resulting density matrices project onto those states in which some particular single-particle states have a definite occupation. The expectation value of some operator Q for these density matrices depends on whether Q involves only the occupations of the single-particle states on which the density matrices project, or others. In the latter case the resulting expectation value is an average over all occupations of the states which do not affect the energy of the transition.

3.2 The Projecting Operator

From the definition of the PDoT in Eq. 1, the integral of the PDoT over all transition energies is the average of squared magnitudes of all matrix elements of the projecting operator. Provided the projecting operator has finite range, its squared magnitude on each initial state is finite, and this is the sum of the squared magnitudes of matrix elements between the given initial state and all final states. Now the average of this finite quantity over all initial states remains finite, and so the total weight of all transitions contributing to a single PDoT is finite. This means that even if the system is infinite, only a finite number of transitions contribute significant weight to the PDoT. Hence, for infinite systems, the choice of projecting operator eliminates almost all transitions from the calculation.

As a result of the preceding argument, projection of transitions onto an operator of finite range is an immensely powerful step in reducing the calculation of properties of extended interacting systems from an infinite to a finite problem. This argument also shows that the choice of projecting operator determines which finite number of transitions contribute to the PDoT or stationary transitions, and hence on which states the density matrices project. Because of the importance of the choice of projecting operator, the remainder of this Subsec. is devoted to discussion of this issue.

The way to think about the projecting operator is as a localized disturbance of the system. The transitions to which this disturbance couples strongly are the ones which contribute significantly to the PDoT, and whose initial and final states have significant weights in the density matrices. Thinking about this intuitively, the transitions to which the disturbance couples strongly resonate or ring for a long time after they are disturbed, while the disturbance dies quickly for the transitions to which it does not couple strongly. This requires strong delta-distributions in the excitation spectra of initial states which couple strongly to the disturbance, and smooth continua in the excitation spectra of weakly coupled initial states. When such spectra are averaged to make the PDoT, the strongest singularities at each energy dominate the average and smooth continua only survive in the average if no states have singular spectra at the same energies. The strength of coupling is just another way of looking at the qualitative differences between the normalizations of transitions, with small normalizations producing strong singularities and large normalizations producing weak singularities or smooth continua.

The simplest example of the relation between the choice of projecting operator and the states in the density matrices is when the projecting operator creates an electron in some localized orbital of a metallic system of interacting electrons. For transitions in which the electron is added at an energy well below the Fermi level, there is a high probability that the orbital is occupied and there is little coupling of the operator to such transitions. Similarly, if the electron is added far above the Fermi level, it rapidly transfers its extra energy to other electrons, and the extra electron disappears into the Fermi sea; the disturbance decays quickly and in this sense couples weakly to such transitions. Only when the electron is added at an energy close to the Fermi energy of the state, does it survive for any length of time. Indeed, when the electron is added to an initial state at its exact Fermi energy, the transition lasts infinitely long because the extra electron cannot lose or gain energy through its interactions with the other electrons. So for each transition energy, it is only the initial state for which that energy equals the Fermi energy that the added electron survives for long times, and the coupling is strong.

In terms on the normalization of transitions, the transition in which the electron goes in at the Fermi level has the least normalization because it requires the least accommodation by the other electrons in the system. If the extra electron goes in above the Fermi energy, then all the other electrons have to acquire a little energy from the new electron in order to equilibrate, and this requires many operators, greatly increasing the normalization of this transition. If the extra electron goes in below the Fermi energy, then the reverse, a hole must be formed to accommodate it by each other electron giving up a little energy, and again the normalization is much larger.

The choice of projecting operator is the choice of how the system is disturbed, and may be viewed as adding a finite number of localized excitations to the system. Initial states and transition energies, for which the interactions immediately dissipate these excitations into an incoherent combination of other excitations, or for which a special combination of existing excitations has to be constructed to accommodate the new excitations, couple weakly to the disturbance so the transitions do not contribute to the PDoT or the polynomial expansion, and the density matrices do not contain projectors for the initial or final states. Only initial states and excitation energies for which the added excitations are long-lived, because they are added at equilibrium, couple strongly to the disturbance. These long lived transitions contribute strongly to the PDoT and polynomial expansion, and these initial and final states dominate the density matrices.

## 3.3 Inner Products

The choice of inner product for the operators also affects the weights of pure transitions in the PDoT and stationary transition, and of initial and final states in the density matrices. One approach to the calculation of thermal properties of interacting system is to replace the inner product in Eqs. 2, 8, and 9 with a Kubo inner product [8, 14]. Viewing the PDoT as an average of excitation spectra over initial states, the Kubo inner product replaces the equal or 'infinite temperature' weighting with a thermal weighting for some specific temperature, so that the resulting PDoT is dominated by the initial state with that specific temperature. Such a thermal PDoT still has a strong singularity at the transition energy for which the excitations are added in equilibrium, but the inner product suppresses the strong singularities at other energies for which the excitations are in equilibrium at different temperatures. Instead, this thermal PDoT includes the less singular contributions from adding excitations out of equilibrium to the state with the specified temperature. The difficulty with inner products such as Kubo's is that they depend on traces over exponentials of the Hamiltonian which is not finite for an extended system. Thermal weighting of the different components in the inner product is equivalent to knowing the equilibrium states, from which expectation values could be calculated directly.

There are ways the inner product in Eqs. 2, 8, and 9 can be changed to alter the weighting of initial states in the PDoT, but without making the inner products more difficult to evaluate. In addition to being as easy to compute as the normalized trace, any new inner product must preserve the orthogonality of non-degenerate stationary states of the Hamiltonian, and this is achieved by simply changing the normalizations of stationary states. The normalized trace in Eq. 4 arises from the usual normalization of stationary states, but many others are possible with the change in normalization effected by a change in the weighting of states in the trace.

New inner products can be constructed using quantities conserved by the Hamiltonian, for example spin, angular momentum, particle number, and so forth. If a quantity is conserved then states with different values of that quantity can be weighted differently in the trace without loss of orthogonality of non-degenerate stationary states, because each of them has a definite value of the conserved quantity. In computing the normalized trace by varying the occupations of different orbitals as in Sec. 2.1, instead of giving equal weight to different values of a conserved quantity, for example occupied and unoccupied or spin up and spin down, these weights can be varied.

4. Simple Examples

Since this work depends on the dominance of transitions with the qualitatively smallest normalizations in the PDoT and polynomial expansion, this Sec. contains two analytic examples of degenerate transitions with different normalizations. In the first example, the system has a localized transition degenerate with a band of extended transitions, and the projecting operator is the annihilation of an electron in a localized combination of the degenerate localized and band states. The second example is of independent electrons hopping on a Bethe lattice, for which there are two bands of states with different localization properties. As a result of the localization properties of the states, there are also two bands of transitions with different localization induced by removing an electron from one site. Only the more localized of these bands contributes to the PDoT.

4.1 A Localized State Degenerate with a Band

This is the most important example because it can be solved analytically and is a paradigm for other systems with degeneracy between transitions having qualitatively different normalizations. A convenient basis for this system of is a set of exponentially localized single-particle orbitals $\{\phi_0, \phi_1, \phi_2, ..., \phi_n, ...\}$ for independent spinless Fermions which evolve according to the Hamiltonian,

$$H = h \sum_{n=1}^{\infty} (c_n^\dagger c_{n+1} + c_{n+1}^\dagger c_n), \tag{17}$$

where $c_n$ annihilates a particle in $\phi_n$. The orbital $\phi_0$ has energy zero and is decoupled from the semi-infinite chain of orbitals $\phi_1, \phi_2, ..., \phi_n, ...$ which each have zero energy but are coupled their neighbors in the chain with matrix element h. The stationary states of this Hamiltonian are products of stationary independent-particle states: $\phi_0$ with energy zero, and a band of delocalized states on the chain, with energies between -2 h and +2 h.

The simplest projecting operator which contains both localized and delocalized transitions is,

$$u_0 = (c_0 + c_1). \tag{18}$$

The Liouvillian **L** for this system commutes the Hamiltonian with operators, so only its action on the annihilation operators for the various orbitals are needed to evolve $u_0$,

$$\mathbf{L} c_n = [H, c_n] = h (c_{n-1} + c_{n+1}), \tag{19}$$

for n greater than zero. Taking the annihilation operators for different orbitals to be orthonormal, the recurrence defined by Eqs. 7-10 is satisfied by,

$$u_{2n-1} = (\sqrt{2}) c_{2n},$$

$$u_{2n} = (\sqrt{2}/2) [(n+1) c_{2n+1} + (c_{2n-1} - c_{2n-3} + \ldots \pm (c_1 - c_0).)]/ \sqrt{[(n+1)(n+2)]},$$

$$b_{2n-1} = h \sqrt{[n/(n+1)]},$$

$$b_{2n} = h \sqrt{[(n+2)/(n+1)]}, \tag{20}$$

for n greater than zero, and $a_n$ is zero for all n.

For energies other than zero the stationary transitions are annihilation of a particle from the band state with that energy,

$$\Psi_\theta = \sum_{n=1}^{\infty} [\sin(n\theta)/\sin\theta] c_n, \tag{21}$$

for $-\pi < \theta < 0$, or $0 < \theta \le \pi$, and the energy of the transition is $2h\cos\theta$. The interesting case is at zero energy where there are two pure transitions, $c_0$ which has a qualitatively small normalization because it is localized, and

$$\Psi_{\pi/2} = \sum_{n=0}^{\infty} (-1)^n c_{2n+1}, \tag{22}$$

which has a qualitatively large normalization because it is delocalized. The expansion of the stationary transitions in the tridiagonal basis $\{u_n\}$, Eq. 12, has coefficients which are the orthogonal polynomials for the weight distribution with half its weight concentrated at zero energy and the other half in a semi-elliptical distribution from -2h to +2h.

The stationary transition of interest is at zero energy where the orthogonal polynomials take the values,

$$P_{2n-1}(0) = 0, \text{ and } P_{2n}(0) = (-1)^n/ \sqrt{[(1+n)(1+n/2)]}. \tag{23}$$

The way that the orthogonal polynomial expansion converges for zero energy shows how the delocalized transition is suppressed. The terms in the expansion up to and including $u_{2N}$ combine to give a coefficient for $c_0$ which is $N/(N+1)$, and so as N goes to infinity, this coefficient converges algebraically to unity. For the other even numbered sites, the coefficient of $c_{2n}$ is always zero for n greater than zero. For the odd numbered sites, the terms in the expansion up to and including $u_{2N}$ combine to give a coefficient for $c_{2n+1}$ which is $(-1)^n/(N+2)$, so individually these components converge algebraically to zero as does the magnitude of their sum.

4.2 Electrons on a Bethe Lattice

The second example is of independent electrons hopping between neighboring sites on a Bethe lattice which has the properties that every site has Z neighbors, and that the only closed paths on the lattice are self-retracing - in other words the lattice has no loops and this is what makes it analytically tractable. Choose one site for the origin of the lattice; label its neighbors 1, 2, ..., Z, and each of their neighbors 11, 12, ..., 1(Z-1), 21, 22, ..., 2(Z-1), ..., Z(Z-1), and so on. A site label $\alpha$ consists of a finite sequence $\alpha_1, \alpha_2, \alpha_3, ..., \alpha_n$ of integers, where $\alpha_1$ is between 1 and Z, and the rest of the integers are each between 1 and Z-1. Each sequence describes how to get from the origin to a particular site by going to the $\alpha_1$ neighbor of the origin, the $\alpha_2$ neighbor of that site, the $\alpha_3$ of the next site, so on, and finally the $\alpha_n$ neighbor of the nth site. The integer $\alpha_2$ and subsequent integers only vary between 1 and Z-1 because the Z neighbors of $\alpha_1, \alpha_2, \alpha_3, ..., \alpha_n$ are the Z-1 sites $\alpha_1, \alpha_2, \alpha_3, ..., \alpha_n, 1$ through $\alpha_1, \alpha_2, \alpha_3, ..., \alpha_n, Z-1$, and $\alpha_1, \alpha_2, \alpha_3, ..., \alpha_{n-1}$. The origin is special because all of its neighbors are in the first shell whereas a site in the nth shell has Z-1 neighbors in the n+1th shell and 1 in the n-1th shell.

The Hamiltonian for independent spinless electrons on this lattice is,

$$H = h \Sigma (c_\alpha^\dagger c_\beta + c_\beta^\dagger c_\alpha) \tag{24}$$

where the sum is over all pairs of neighboring sites $\alpha$ and $\beta$ on the lattice, and $c_\alpha$ annihilates the electrons on site $\alpha$. This Hamiltonian has extended states, like Bloch states, for which the electron has equal probability of being on each site. At energy Z h there is the state with coefficient 1 for each site, and this band extends to - Z h where the sign of the coefficient alternates between neighbors but has constant magnitude. Between - Z h and Z h are states whose coefficients on neighboring sites have the same magnitude, but differ in phase by angles between zero (Z h) and $\pi$ (- Z h). The Hamiltonian in Eq. 24 also has states which are exponentially localized around one site which can be taken to be the origin. These localized states form a band with energies ranging from $-2h\sqrt{(Z-1)}$ to $2h\sqrt{(Z-1)}$, and taking $\theta$ to be a parameter between 0 and $\pi$, the coefficients for sites on the nth shell of neighbors of the origin is $Z^{-1}(Z-1)^{1-n/2}\sin[(n+1)\theta]/\sin\theta$, with unit coefficient on the origin. Note that these localized states break the symmetry of the lattice, are purely real, and for Z greater than two, the amplitudes of these states decreases exponentially on each succeeding shell of neighbors of the origin.

Taking the annihilation of an electron at the origin as the local disturbance which initiates the recursion,

$$u_0 = \sqrt{2} \, c. \tag{25}$$

Both localized and delocalized states have non-zero amplitude on the origin, so this disturbance could remove an electron from either kind of state. Proceeding with the recursion gives,

$$u_n = \sqrt{2} \, Z^{-1/2} (Z-1)^{(1-n)/2} \, \Sigma \, c_\alpha, \tag{26}$$

where the sum is over annihilation operators for electrons on all nth neighbors of the origin, and

$$a_n = 0 \text{ for all } n, \; b_1 = h\sqrt{Z}, \text{ and } b_n = h\sqrt{(Z-1)} \text{ for n greater than 1.} \tag{27}$$

The PDoT for the Bethe lattice is proportional to the imaginary part of the continued fraction, Eq. 5, with the above parameters substituted into it. Because the parameters are constant for n greater than one, the continued fraction can be evaluated analytically to give for E real and Z greater than 2,

$$R(E) = (1/2)\{(2-Z)E \pm Z\sqrt{[E^2 - 4(Z-1)h^2]}\}/(E^2 - Z^2 h^2), \tag{28}$$

where the positive sign applies when E is $|2h\sqrt{(Z-1)}|$ or greater, the negative sign applies when E is $-|2h\sqrt{(Z-1)}|$ or less, and the continued fraction has two branches between $-|2h\sqrt{(Z-1)}|$ and $|2h\sqrt{(Z-1)}|$. Since the PDoT is proportional to the imaginary part of the continued fraction, it is non zero between $-|2h\sqrt{(Z-1)}|$ and $|2h\sqrt{(Z-1)}|$, corresponding to the band of localized transitions. The zeros in the denominator at h Z and -h Z suggest that the PDoT might also have delta-distributions at those energies which correspond to the edges of the band of delocalized transitions, however, substituting h Z and -h Z into the numerator of Eq. 28 gives zero, so there are no delta-distributions in the PDoT.

In this example, only the localized transitions contribute to the PDoT and the delocalized transitions are completely absent, even at energies where there are no localized transitions to suppress them. This is different from the first example where the only delocalized transition absent from the PDoT is the one which is degenerate with the localized transition. The difference between the examples is that on the Bethe lattice the normalizations of the two kinds on transitions diverge exponentially from one another, while for the chain and localized state, the two kinds of transitions only diverge algebraically.

5. The Heisenberg Chain

The ground state energy of an anti-ferromagnetic spin one-half Heisenberg chain is known for the infinite chain from analytic calculations [15] and correlation functions have been calculated numerically [16] for finite chains of up to about 30 spins. The purpose of the calculations presented here is to demonstrate numerically the use of the recursion method from Sec. 2 to calculate the PDoT, and the interpretation of the PDoT from Sec. 3 in computing the properties of interacting electrons. Although these results have been previously obtained analytically, this numerical calculation of the binding energy and zone boundary spin wave energies for states of an infinite, spin 1/2 Heisenberg chain is new.

5.1 The Projecting Operator

The first step in this calculation is the choice of a projecting operator which is short ranged in that it only changes the occupations of a few localized single particle orbitals. While the binding energy of an infinite Heisenberg chain is a global and hence infinite quantity, what is finite is the binding per spin of the chain in its various states. This binding energy per spin for the infinite chain is simply the energy required to remove the last spin from a semi-infinite chain, a short ranged disturbance which avoids the numerical and termination errors in trying to calculate the limit of the binding energy per spin for long but finite chains.

In the Heisenberg chain each electron is spatially localized, but can have either spin. It is convenient to define operators $c_{n\uparrow}$ and $c_{n\downarrow}$ which respectively annihilate the electron from the spin up and spin down orbitals on the nth site, where the sites of the semi-infinite chain are numbered 0, 1, 2, ... The energy of removing a spin from the semi-infinite chain is related to the projecting operator which annihilates an electron from either the up or down spin orbital of site 0, and since no direction is preferred, it doesn't matter which operator is used, so define,

$$u_0 = \sqrt{2}\, c_{0\uparrow} \tag{29}$$

where the root of two normalizes the operator. In the space of states where each site is occupied by a single electron, the Hamiltonian for the Heisenberg chain is,

$$H = J \sum_{n=0}^{\infty} \sum_{\sigma=\uparrow\downarrow} (c_{n\sigma}^\dagger c_{n+1\sigma}^\dagger c_{n+1\sigma} c_{n\sigma} + c_{n-\sigma}^\dagger c_{n+1\sigma}^\dagger c_{n+1-\sigma} c_{n\sigma}) \tag{30}$$

and the Liouvillian is the commutator of H.

5.2 Tridiagonalization of the Liouvillian

The calculation proceeds by commuting H with $u_0$, then projecting out the component of $u_0$ to determine $a_0$, normalizing the remainder to determine $b_1$, and so on according the Eqs. 8-10. The projection and normalization depend on the choice of inner product for operators, which in this case is the normalized trace inner product defined in Eq. 4, with the trace taken over all states in which the sites are singly occupied. Note that this illustrates how the choice of inner product determines which transitions contribute to the PDoT, in this case the transitions between states in which each site on the chain is single occupied. The trace used here is restricted to a subspace of all electronic states, but the Liouvillian remains Hermitian with respect to this inner product as can easily be verified.

While the tridiagonalization can be done by hand for a number of levels, it is convenient to do it numerically using a program developed by W.M.C. Foulkes, and modified for this calculation. The tridiagonal matrix elements for the first thirteen operators of the recursion basis are presented in Table 1. While both diagonal and off-diagonal elements fluctuate from level to level, there also seem to be simple trends. This behavior is better seen in Figs. 1 and 2 where fluctuations in both kinds of parameter seem to be decreasing with increasing n; to a constant for the diagonal elements as shown by the linear plot and a constant times $\sqrt{n}$ for the off-diagonal elements as shown by the log-log plot.

If the $a_n$ were actually constant and the $b_n$ proportional to $\sqrt{n}$, the PDoT would be a Gaussian [17], as can be seen from the recurrence relation for Hermite polynomials which are orthogonal with respect to a Gaussian weight distribution. Changing a few of the initial matrix elements from their ideal values produces smooth variations of the PDoT about a Gaussian, and with increasing numbers of matrix elements differing from their ideal values the deviations from the Gaussian become sharper. If the matrix elements only converge slowly to the ideal values for the Gaussian, then there can be sharp structure on top of the underlying Gaussian.

The apparent convergence of the tridiagonal matrix elements in Table 1 to those of a Gaussian suggests that the PDoT generated by removing an electron from the end of a Heisenberg chain consists of a Gaussian with additional structure superimposed on it. The Gaussian makes sense in a statistical approximation for which the PDoT is the probability distribution in energy for transitions made up of independent spin flips of individual electrons caused by the removal of the end electron. In this approximation there is a probability distribution in energy for the transition of each electron, and the probability distribution for the whole transition is simple the convolution of the distributions for each electron. In this approximation the PDoT is the distribution of a sum of independently distributed variables, so it is Gaussian by the central limit theorem. Taking the probability distribution for a single spin flip to have width about 2J is consistent with calculated tridiagonal matrix elements. In terms of this statistical approximation, the deviations of the tridiagonal matrix elements from their values for a Gaussian arise from correlations between the different electronic transitions which is due to their interactions.

5.3 The Difference Between Logarithms of PDoTs

The PDoT for the Heisenberg chain is shown in Fig. 3. This Fig. was calculated from Eq. 5 using the computed parameters in Table 1 followed by terminating parameters $a_n$ of -0.25 and $b_n$ given by $0.7698\, n^{0.63568}$, for n greater than 12. The energy dependence of this terminator was retained to n equal 8,000, beyond which E was fixed at -0.25. The part of the terminator beyond n equals 8,000 can be evaluated analytically for this particular value of E and then the rest of the continued fraction is evaluated numerically.

Although this PDoT is distinctly non-Gaussian, it is difficult to see the additional structure. In order to make clearer the structure in the PDoT due to electron correlations, the calculated PDoT is compared with an ideal PDoT obtained by continuing the same $a_n$ and $b_n$ as the terminator to the top of the fraction. This comparison is shown in Fig. 4 where the difference between the logarithms of the computed and ideal PDoTs is plotted against energy. Logarithms are used to show structure where both PDoTs get very small and to make decreases in the PDoT comparable with increases.

There are many different ways this PDoT could be presented including quadrature methods [18] and other terminators. Quadrature has the advantage of never introducing spurious structure and the disadvantage of poor resolution of what structure there is. When applied to the parameters in Table 1., quadrature clearly shows the peaks near -1.0 and 3.0 as well as minima near -2, 0.4, and 3.5 in Fig. 4, indicating that this is real structure in the PDoT. Termination methods such as those used in Figs. 3 and 4 have the possibility of better resolution than quadrature, but they can also introduce spurious structure, so it is important to compare the PDoTs obtained by different methods. Of the many different terminators tried for this problem, the terminator used to produce Fig. 4 gives the best resolution of the features seen in quadrature without introducing any spurious features between -1.0 and 3.0. This terminator also partially resolves features near -7, -2, 4, and 7, although without confirmation from quadrature, the significance of these later features is in doubt.

5.4  Comparison with Analytic Results and Numerical Calculations for Finite Rings

The choice of a Heisenberg chain as an example is motivated by the analytic results which exist for this system and can be used for comparison with the calculations presented here. Table 2 contains a selection of important energies of the infinite, spin 1/2, Heisenberg chain in units of J for the Hamiltonian in Eq. 36, which have been derived from previous analytic work [19]. In the absence of other numerical approaches for the infinite Heisenberg chain, results from the above calculation can be compared to extrapolations of numerical results for finite rings of spins.

From Sec. 2, the interpretation of the PDoT in Fig 3 is as the probability distribution for transitions of the semi-infinite Heisenberg chain, induced by annihilating a spin up electron from the first site on the chain. From Sec. 3.1, the transition contributing to the PDoT at each energy is the most localized transition with that energy. The most localized transition possible is one in which the removal of the spin up electron does not affect any of the other electrons. That transition is simply the removal of the end electron from the state in which all electrons have up spin, a ferromagnetic state. Both the state with and without the end electrons are stationary states of the Hamiltonian and the transition is just the operator which takes the system from one state to the other. Such a localized transition should produce a strong peak in the PDoT at -1, because the total energy of the chain is reduced by the energy of one ferromagnetic bond. Indeed there is such a peak in the PDoT and it is even better resolved in Fig. 4.

There are other states of the chain which are almost ferromagnetic and so have energies of almost J per electron. If the end electron is removed from one of these states at exactly the same energy as the energy per electron of the state, then there is no energy left over to make a spin wave or other excitation, nor is any energy needed from a spin wave to make the transition. Such transitions are less localized than that for the ferromagnetic state but the more ferromagnetic the state, the more localized the transition because there are fewer spin flips to propagate down the chain. Transitions in which the end electron is removed at exactly the energy per electron of the state form a band because they vary smoothly from the ferromagnetic state to the opposite extreme, the antiferromagnetic state.

Removal of the end electron from the antiferromagnetic state is the least localized transition which does not involve absorbing or emitting a spin wave. The antiferromagnetic state consists of subtly correlated spin flips between neighboring spins and these correlations vary with distance from the end of the chain. So when the end electron is removed, all the correlations in the chain have to readjust because the end of the chain is one site nearer. This produces a transition which extends over the whole chain, only decreasing weakly with distance from the end. In the band of transitions which do not involve spin waves, this transition between antiferromagnetic states should have the smallest value of the PDoT, and indeed there is a minimum at the transition energy for removing one electron from the anti-ferromagnetic state, at

2 Ln 2 - 1, about 0.3863.

Heisenberg Hamiltonians have been diagonalized for finite rings of N spins, and the errors in the binding per electron of the antiferromagnetic state relative to the ferromagnetic state are found [19] to be well approximated by $\alpha/N^2$ where $\alpha$ is about 2.5 for N odd and about half that for N even. This illustrates the general difficulty of extrapolating from calculations for finite systems, namely that their convergence is at best power-law in contrast with continued fractions for the infinite system which converge exponentially to the precision of the arithmetic used [11]. Varying the number of levels in the continued fraction for the infinite Heisenberg chain show that features in the PDoT converge at a rate of about one decade for every 8 levels. This is consistent with an error of about 2.5% in the relative binding energies of the ferromagnetic and antiferromagnetic states in Fig. 4 which was calculated with 13 levels. Furthermore, this estimate of the convergence rate predicts that the accuracy of the continued fraction will overtake that of the largest finite diagonalization currently possible (32 spins) at about 24 levels which is well within the capabilities of methods using variable precision arithmetic[20], though not within the capabilities of the fixed precision program used in this work.

This band of transitions without spin waves must have the ferromagnetic and antiferromagnetic binding energies as extrema because beyond these binding energies there can be no transitions without spinwaves. The qualitative nature of the transitions change at these band edges and so the PDoT has singularities, like the Van Hove singularities at the edges of other kinds of bands. These singularities are sharp and while the numerical calculations presented here do not resolve the nature of these singularities, they do produce the sharp features in Fig. 4.

According to the discussion in Sec. 3.1, contributions to the PDoT in Fig. 4 at energies below -1 or above 2 Ln 2 - 1 must come from the transitions, having the qualitatively smallest normalizations, in which spin waves are either emitted or absorbed. One might think that these would be the transitions in which the fewest spin waves are emitted or absorbed, but because the degeneracy of transitions increases with the number of spin waves involved, transitions of qualitatively similar normalization can be constructed for any number of spin waves. The spin wave spectra are one-dimensional, so they have van Hove singularities only at the center and boundary of each Brillouin zone, and the spectrum for removal of the electron only has singularities at the ferromagnetic and antiferromagnetic states, so the strongest singularities occur when the contribution from the electron and each spin wave is singular. These strong singularities occur in the PDoT only when some number of zone boundary spin waves are absorbed or emitted from either the ferromagnetic or antiferromagnetic states. Note that adding a spin wave to the ferromagnetic state reduces the energy of the state when J is taken to be positive

It follows from the above that there should be singularities in the PDoT at the energies for the reversible removal of the end electron with no spin waves absorbed or emitted from the ferromagnetic state, -1, and antiferromagnetic state, 2 Ln 2 - 1, and at these energies plus or minus the energies of integer numbers of zone boundary spin waves. The transition in which the end electron is removed from the ferromagnetic state is strongly localized, so there should be strong singularities in the PDoT at this energy plus or minus the energies of integer numbers of ferromagnetic zone boundary spin waves. These transition energies are 3, and 7 for the absorption of one and two spin waves from the ferromagnetic state; and at -5 and -9 for emission of one and two spin waves. There is a strong, resolved peak close to 3 in Fig. 4 and a less well resolved peak close to 7, which could be singularities for absorption of ferromagnetic zone boundary spin waves. There is a strong peak at -7 which might correspond to emission of one spin wave, but -9 is outside the range of this calculation. Similarly, we might expect to find minima in Fig. 4 for the energies at which zone boundary spin waves are emitted or absorbed from the antiferromagnetic state. These should occur at 2 Ln 2 - 1 $\pm\pi$ or $\pm 2\pi$, which are approximately -5.9, -2.8, 3.5, and 6.7. Figure 4 shows a well resolved minimum near 3.5 and a not so well resolved minimum near -2; -5.9 and 6.7 are both close to the peaks expected for emission and absorption of two ferromagnetic spin waves, so it is not clear what should be seen at this resolution.

The correspondence between features in Fig. 4 and singular transitions of the infinite Heisenberg chain seems more than accidental, although there must be some doubt until the PDoT for this system is calculated to higher resolution.

6. Localization, Thermodynamics, and Statistical Mechanics

This paper presents a method for calculating the energies of transitions between states of interacting electrons because, unlike the energies of the states themselves, the energies of transitions do not diverge with increasing size of the system. However, like states, the number of transitions increases exponentially with system size, and the problem of finding a particular transition is just as difficult as that of finding a particular state. In this sense, the problem of calculating properties of interacting systems is the same as that of finding a needle in an exponentially large haystack.

While finding a needle in a haystack usually requires the long process of examining each stalk of hay, the methods developed above find the properties of equilibrating systems in a time independent of the system size. The efficiency of these methods is due to equilibration in which energy or other physical quantities are distributed homogeneously throughout the system rather than being stuck where they are deposited. It is equilibration which makes most transitions irreversible by distributing a localized disturbance throughout the entire system, and makes the reversible transitions singular. In its simplest form, the critical difference is between equilibrating systems in which most transitions are irreversible carrying currents of energy or other quantities to infinity, and non-equilibrating systems in which most transitions are reversible and do not carry currents because they are localized. Put more abstractly, it is the difference between systems in which most of the transitions belong to time-reversal doublets, carry currents, equilibrate, and are therefore irreversible; and systems in which most of the transitions belong to time-reversal singlets, do not carry currents, do not equilibrate, and are therefore reversible.

Since the method presented in this paper yields equilibrium properties of interacting electrons, there must be connections between this method and both thermodynamics and statistical mechanics. In what follows, two of these connections are outlined. The first is to thermodynamics by means of a proposed relation between the normalization of transitions and changes in the entropy of quantum states. The second is to statistical mechanics by interpreting the continued fraction and polynomial expansions as high-temperature expansions for the canonical ensemble.

6.1 Entropy and the Normalization of States

Classically, entropy is defined as the logarithm of the phase space volume explored as a state evolves. Since evolution is quantum mechanical on microscopic scales, there must be some connection between entropy and quantum states, at least in the limit of large quantum numbers. If the entropy of a quantum state is to correspond to the entropy of a classical state in the limit of large quantum numbers, then the entropy of a quantum state must be defined as the logarithm of a quantum mechanical quantity which becomes the phase space volume in the classical limit. This is where the notion of localized orbitals and localized operators, on which the properties of the PDoT depend, become important again.

The probability distribution of a localized orbital decreases exponentially with distance from its center in both direct space and momentum space, and a localized operator simple changes the occupation of a few localized orbitals. Because the orbital is exponentially localized both in direct and momentum space, it is also exponentially localized in classical phase space, and while it has a finite size with respect to microscopic length and momentum scales, it approaches a point in the limit of classical length and momentum scales. However, one point or even a finite number of points in classical phase space do not make a volume, so this classical limit only produces a non-zero entropy when the number of points is infinite which requires either an infinite quantum system or the limit of infinite quantum number for a finite system. Since the classical entropy is the logarithm of the volume, differences in the entropies of quantum states are independent of the volume of each localized orbital.

Because the number of localized orbitals spanned by a quantum state is counted by its normalization, the above argument implies that for infinite systems the entropy associated with a quantum state is related to the way its normalization goes to infinity as the system becomes large. While there are many ways the divergence of state normalizations could be compared, a convenient one is to compare them to the number of localized, single-particle orbitals - essentially the volume of the system. The localized orbitals which span the states of interacting systems are products of localized single-particle orbitals, so the normalization of the state can vary from being exponential in the number of single-particle orbitals to being independent of their number. Given n single-particle orbitals $\{\phi_1, \phi_2, \phi_3, ...\phi_n\}$ and some product orbital $\Phi$ (this product can include any single-particle orbital, not just $\phi_1, \phi_2, \phi_3, ...\phi_n$) which contributes significantly to the state $\Psi$, $(\Psi,\Psi)_n$ is the normalization of that part of $\Psi$ generated by varying the occupations of $\{\phi_1, \phi_2, \phi_3, ...,\phi_n\}$ in $\Phi$. This measure of entropy is then,

$$S[\Psi] = \lim_{n\to\infty} Ln\ (\Psi, \Psi)_n / Ln\ n, \qquad (31)$$

where the limit is taken as the occupations of all single-particle orbitals are allowed to vary, thus obtaining the normalization of the entire state. The properties of localized orbitals make this entropy independent of the choice of orbitals including the choice of $\Phi$ provided its coefficient in $\Psi$ is not zero.

The variation of occupation numbers in the normalization of a quantum state brings the discussion back to the localization properties of transitions. Each localized operator in the expansion of a transition, Eq. 12, changes the occupations of a finite number of single-particle orbitals and so, when applied to a state, the local operator cannot increase its normalization, although it might reduce it by annihilating orbitals. A finite sum of local operators, can at most multiply the normalization of a state by a finite number which, according Eq. 31, does not change its quantum entropy. In order to change the quantum entropy of a state, a transition operator must have a number of terms comparable to the number of single-particle orbitals, and in that sense be delocalized. It seems that the identification of delocalization with irreversible, and therefore entropy increasing, transitions is at least consistent with relating the entropy of a state to the divergence of its normalization.

6.2 High-Temperature Expansions

In Sec. 2 the continued fraction expansion for the PDoT and the polynomial expansions for the density matrices are derived from the dynamics of the interacting electrons. However, the same results can be viewed as high-temperature expansions obtained from the canonical ensemble. The connection between these two views is the choice of inner product , Eq. 4. In a dynamical approach this inner product is simply a convenience for extracting numerical parameters from powers of the Liouvillian. In statistical mechanics, the inner product defines how states are weighted in the calculating the tridiagonal matrix elements which are expectation values for some canonical ensemble. For the calculations presented in Secs. 4 and 5, it is convenient to use weights for canonical ensembles at infinite temperature.

High-temperature expansions of the partition function and free energies have been widely used to estimate critical temperatures and critical exponents by asymptotic analysis of the coefficients for power series in inverse temperature. However, such analyses have been problematic because the series diverge at critical temperatures and only a finite number of terms can be computed. From this perspective, the inner product used in this work leads to expansions of the PDoT and density matrices for infinite temperature, but they are expanded in energy rather than inverse temperature. The advantage of the approach adopted here is that the continued fraction expansion of the PDoT and the polynomial expansions of the density matrices converge exponentially. Because the expansions used in this work converge so strongly, they can be used to calculate the properties of all states including the ground state, not just the high-temperature phase as is on expansions. This has been demonstrated in practice for the Heisenberg spin chain, and in principle, for other interacting systems.

Acknowledgments

The Author acknowledges with gratitude, permission from Matthew Foulkes to modify and use an unpublished program for tridiagonalizing the Liouvillian for a Heisenberg spin chain. Useful discussions with Matthew Foulkes, James Annett, Guna Rajagopal, and Wolfram Arnold are also acknowledged. Some of the numerical calculations were performed on computers provided by the National Science Foundation's Office of Science and Technology Infrastructure under grant no. STI-9413532, and other numerical calculations were performed on the Hitachi S3600 located at the University of Cambridge High Performance Computing Facility.

**Figures**

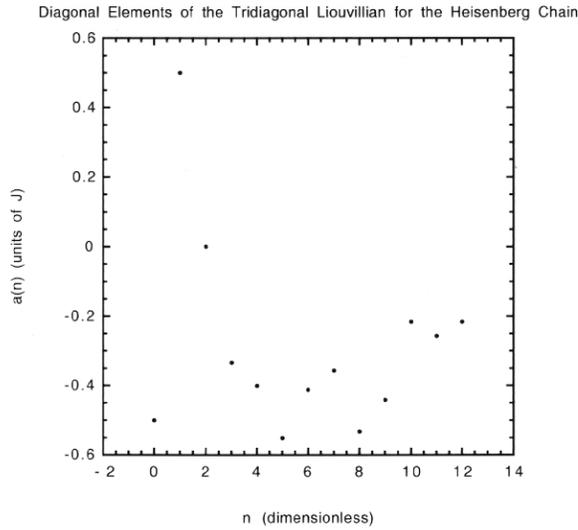

Figure 1. The $a_n$ obtained by removing the end electron of a semi-infinite Heisenberg chain.

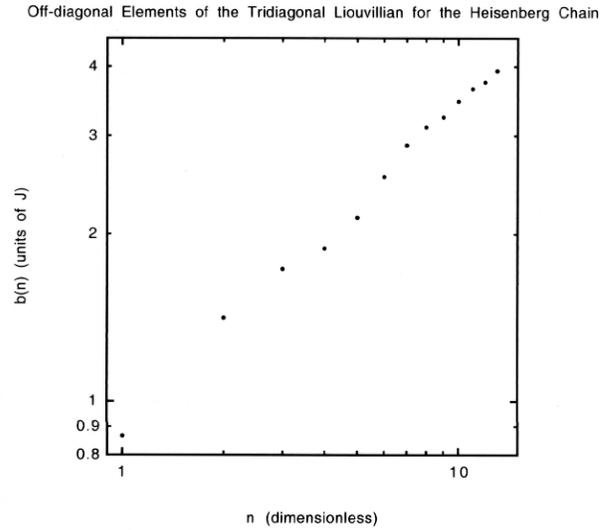

Figure 2. The logarithm of $b_n$ plotted against the logarithm of n for removing the end electron from a semi-infinite Heisenberg chain.

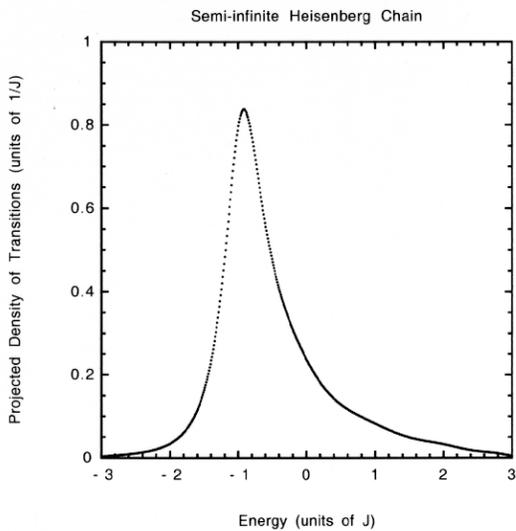

Figure 3. The projected density of transitions for removing a spin up electron from the end of a semi-infinite spin chain, calculated using a Gaussian terminator.

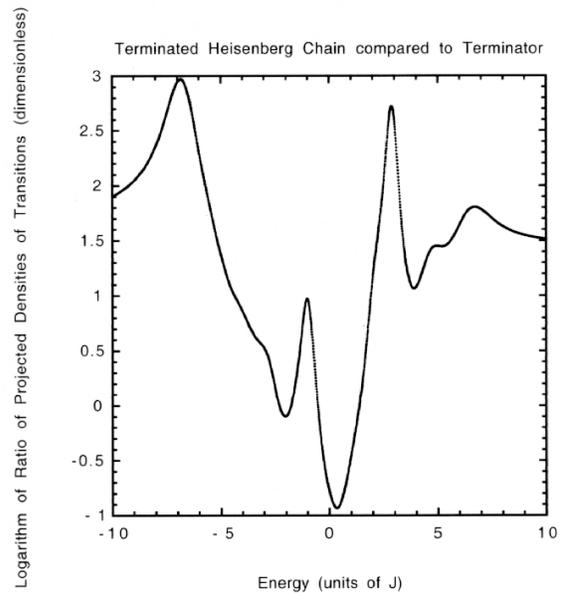

Figure 4. The difference between logarithms of the terminated PDoT for the semi-infinite Heisenberg chain and the PDoT of the terminator.

**Table Captions**

Table 1. Tridiagonal matrix-elements for the first fourteen operators of the recursion basis, Eq. 7.

Table 2. Microscopic energies in units of J for the Heisenberg spin chain in Eq. 30, derived from [19].

**Table 1**

| | |
|---|---|
| $a_0 = -0.50000000$ | $b_1 = 0.8660254$ |
| $a_1 = 0.50000000$ | $b_2 = 1.4142136$ |
| $a_2 = 0.00000000$ | $b_3 = 1.7320508$ |
| $a_3 = -0.33333333$ | $b_4 = 1.8856181$ |
| $a_4 = -0.40104167$ | $b_5 = 2.1414232$ |
| $a_5 = -0.55120238$ | $b_6 = 2.5331972$ |
| $a_6 = -0.41212965$ | $b_7 = 2.8867582$ |
| $a_7 = -0.35662244$ | $b_8 = 3.1121898$ |
| $a_8 = -0.53222315$ | $b_9 = 3.2451531$ |
| $a_9 = -0.44093913$ | $b_{10} = 3.4638295$ |
| $a_{10} = -0.21551313$ | $b_{11} = 3.6457083$ |
| $a_{11} = -0.25669094$ | $b_{12} = 3.7509117$ |
| $a_{12} = -0.21553442$ | $b_{13} = 3.9309091$ |

**Table 2**

| | |
|---|---|
| Energy per electron in the ferromagnetic state | 1 |
| Energy per electron in the antiferromagnetic state | 1 - 2 Ln 2 |
| Energy of a zone boundary spin wave in the ferromagnetic state | -4 |
| Energy of a zone boundary spin wave in the antiferromagnetic state | π |